\documentclass[12pt]{iopart}

\usepackage{graphicx}
\usepackage{multirow}
\usepackage{ulem}
\usepackage{color}
\usepackage{iopams}

\newcommand{\opcdag}{\ensuremath{\hat{c}^\dagger}}
\newcommand{\opc}{\ensuremath{\hat{c}}}

\newcommand{\opadag}{\ensuremath{\hat{a}^\dagger}}
\newcommand{\opa}{\ensuremath{\hat{a}}}

\newcommand{\Norb}{\ensuremath{{N_\mathrm{orb}}}}

\newcommand{\bW}{\ensuremath{\boldsymbol{W}}}

\begin{document}
\title{Hybridization expansion Monte Carlo simulation of multi-orbital quantum impurity problems: matrix product formalism and improved Monte Carlo sampling}

\author{Hiroshi Shinaoka$^1$}
\author{Michele Dolfi$^1$}
\author{Matthias Troyer$^1$}
\author{Philipp Werner$^2$}
\address{$^1$Theoretische Physik, ETH Z\"{u}rich, 8093 Z\"{u}rich, Switzerland}
\address{$^2$Department of Physics, University of Fribourg, 1700 Fribourg, Switzerland}

\date{\today}

\begin{abstract}
We explore two complementary modifications of the hybridization-expansion continuous-time Monte Carlo method, aiming
at large multi-orbital quantum impurity problems.
One idea is to compute the imaginary-time propagation using a matrix product states representation.
We show that bond dimensions considerably smaller than the dimension of the Hilbert space are sufficient to obtain accurate results, and that
this approach scales polynomially, rather than exponentially with the number of orbitals. 
Based on scaling analyses, we conclude that a matrix product state implementation will outperform the exact-diagonalization based method 
for quantum impurity problems with more than 12 orbitals.
The second idea is an improved Monte Carlo sampling scheme which is applicable to all variants of the hybridization expansion method.
We show that this so-called sliding window sampling scheme speeds up the simulation by at least an order of magnitude for a broad range of model parameters, with the largest improvements at low temperature.
\end{abstract}

\maketitle

\section{Introduction}
Quantum impurity models appear in various contexts in condensed matter physics. 
An important example is the dynamical mean-field theory (DMFT)~\cite{Georges96} for strongly correlated electron systems.
In a DMFT calculation, a correlated lattice model is mapped to an impurity problem whose bath degrees of freedom are self-consistently determined. 
Although the DMFT formalism was originally proposed for the single-band Hubbard model, it can be extended to multi-orbital systems and cluster-type impurities.~\cite{Maier-review}
Furthermore, DMFT can be combined with density functional theory based ab-initio calculations, to describe strongly correlated materials such as transition metal oxides.~\cite{Kotliar-review}
For these applications, it is important to develop efficient algorithms to solve quantum impurity problems with multiple orbitals or sites.

In recent years, two complementary types of continuous-time quantum Monte Carlo (MC) impurity solvers have been developed, which are based on a stochastic sampling of perturbation expansions:
the weak-coupling method~\cite{Rubtsov05} and the hybridization expansion method.~\cite{Philipp06a,Philipp06b}
The former approach is based on a perturbation expansion in powers of the Coulomb interaction terms,  
while the latter one treats the local Coulomb interactions exactly and instead expands the partition function in the coupling between the impurity and the bath.
For describing strongly correlated materials,
the latter approach is typically favored because of its ability to treat general interactions such as spin flips, and because the average perturbation order of the hybridization expansion
is relatively low in the strongly correlated regime. 
The algorithm was further extended to treat retarded interactions,~\cite{Philipp10} which has recently been used in a extended DMFT study of the effects of long-range interactions.~\cite{Ayral13} 

A drawback of the hybridization expansion approach is that the computational effort scales exponentially with the number of sites or orbitals, because the dimension of the Hilbert space grows exponentially.
Without additional approximations, this limits the application to small impurity models with up to five orbitals, even if one uses an implementation based on sparse-matrix exact-diagonalization techniques.~\cite{Lauchi09}

On the other hand, various wavefuction based theories have been developed for interacting fermionic lattice models.
In particular, the ground states of one-dimensional (1D) systems can be described essentially exactly by the formalism of matrix product states (MPS)~\cite{Ostlund95} with reasonable computational effort. The MPS formalism is known to be equivalent to the density matrix renormalization group (DMRG).~\cite{White92a,White92b}
It has also been used to solve impurity problems.~\cite{Nishimoto04,Carsten04,Carsten05,Nishimoto06,Robert11}
In such MPS based calculations, the bath is represented by a 1D chain (or 1D chains) attached to the impurity, which results in an exponential growth of the computational cost with the number of sites or orbitals in the impurity.
Furthermore, it is not trivial to extend the formalism to a non-diagonal coupling between the impurity and the bath, or to retarded interactions.

A possible direction for the development of flexible impurity solvers for large multi-orbital systems may be to combine these two approaches, i.e., the hybridization expansion and the MPS formalism.
In this paper, we propose and test such a combined approach, in which the local interaction is treated using an MPS representation. More specifically, we perform the imaginary time evolution, which is given by the local impurity Hamiltonian, using the MPS formalism. We test the accuracy of the imaginary time evolution and compare its performance with that of the exact approach using a sparse-matrix exact-diagonalization technique.

Another direction of research is to develop a more efficient MC sampling algorithm.
For the continuous-time MC method based on the hybridization expansion,
one stochastically samples configurations represented by creation and annihilation operators of the local degree of freedoms on the imaginary time interval.
In estimating the weight of a configuration, the most costly part in multiorbital cases is evaluating the trace of a matrix product over the local degrees of freedom of the quantum impurity.
This matrix product consists of imaginary-time evolution operators as well as creation and annihilation operators.
The cost of evaluating the trace grows as temperatures is lowered, because the expansion order increases. 

The trace can be evaluated either by the matrix formalism,~\cite{Philipp06b,Haule07} by sparse-matrix exact-diagonalization techniques (Krylov method)~\cite{Lauchi09} or by an MPS version of the Krylov method. 
In the former formalism, all operators are represented by matrices in the eigenbasis of the local Hamiltonian,
and the matrix product is computed by multiplying the matrices one by one.
In the latter formalism, the trace is computed by performing the imaginary-time evolution starting from eigenstates using 
the basis in which operators are represented as sparse matrices.
In this paper, we call this the Krylov method or Krylov-sparse-matrix method.
It was shown that the Krylov method is superior in performance for impurity problems involving more than 4 orbitals as local degrees of freedom.~\cite{Lauchi09}

For the matrix formalism, an efficient MC sampling scheme based on a tree structure has been proposed to suppress the growth of the computational cost at low temperatures.~\cite{Emanuel-review}
Instead of recomputing the matrix product from scratch at each MC step, one reuses  partial products of matrices that have been previously computed and stored.
By using a tree data structure, the cost can then be reduced from $O(\beta)$ to $O(\log \beta)$, where $\beta$ is the inverse temperature.
However, these ideas based on storing matrix products cannot be applied to the Krylov method.
Thus, an alternative efficient MC sampling algorithm needs to be developed for the Krylov method.

The rest of the paper is organized as follows.
In Sec.~\ref{sec:hyb}, we describe the hybridization expansion algorithm.
The Krylov method is described in Sec.~\ref{sec:krylov}.
The quantum impurity models used for the present study are defined in Sec.~\ref{sec:model}.
In Sec.~\ref{sec:krylov-mps}, we propose a combined approach of the Krylov method and the matrix-product formalism.
We propose an improved MC sampling algorithm for the Krylov method in Sec.~\ref{sec:improved-sampling}.
A summary is given in Sec.~\ref{sec:summary}

\section{Hybridization expansion algorithm}\label{sec:hyb}
A fermionic quantum impurity model is defined by the following Hamiltonian:
\begin{eqnarray}
  \mathcal{H} &=& \mathcal{H}_\mathrm{loc} + \mathcal{H}_\mathrm{mix} + \mathcal{H}_\mathrm{bath},
\end{eqnarray}
where 
\begin{eqnarray}
  \mathcal{H}_\mathrm{loc} &=& \sum_{\alpha,\beta} t_{\alpha,\beta} \opcdag_\alpha \opc_\beta + \sum_{\alpha,\beta,\gamma,\delta} U^{\alpha,\beta,\gamma,\delta} \opcdag_\alpha \opcdag_\beta \opc_\gamma \opc_\delta,\\
  \mathcal{H}_\mathrm{bath} &=& \sum_{k,\alpha} \epsilon_{k,\alpha} \opadag_{k,\alpha} \opa_{k,\alpha},\\
  \mathcal{H}_\mathrm{mix} &=& \sum_{k,\alpha,\beta} V_k^{\alpha,\beta} \opadag_{k,\alpha} \opc_\beta + \mathrm{h.c.}.
\end{eqnarray}
The term $\mathcal{H}_\mathrm{loc}$ describes an impurity with chemical potentials, intra-orbital hoppings and two-body interactions,
where $\alpha$ and $\beta$ are combined orbital and spin indices. (We call the combined index of spin and orbital a flavor.)
$\mathcal{H}_\mathrm{bath}$ describes a non-interacting bath with quantum numbers $k$ and spin/orbital index $\alpha$.
The hybridization term $\mathcal{H}_\mathrm{mix}$ describes the exchange of electrons between the impurity and the bath.

In the hybridization expansion impurity solver, one expands the partition function $Z = \mathrm{Tr} \left[ e^{-\beta \mathcal{H}}\right]$ with respect to the hybridization term $\mathcal{H}_\mathrm{mix}$ as
\begin{eqnarray}
  Z &=& \mathrm{Tr} \left[ e^{-\beta \mathcal{H}}\right]\nonumber \\
  &=& \mathrm{Tr} \left[ e^{-\beta \mathcal{H}_1} T e^{-\int_0^\beta \mathrm{d} \tau \mathcal{H}_2(\tau) }\right]\nonumber \\
  &=& \sum_{n=0}^\infty \int_0^\beta \mathrm{d} \tau_1 \cdots \int_{\tau_{n-1}}^\beta \mathrm{d} \tau_n  (-1)^n\nonumber \nonumber\\
  && \hspace{20mm}\times \mathrm{Tr}
  \left[ 
  e^{-(\beta-\tau_n) \mathcal{H}_1}
  \mathcal{H}_2 
  e^{-(\tau_n-\tau_{n-1}) \mathcal{H}_1}
  \cdots
  \mathcal{H}_2 
  e^{-\tau_1 \mathcal{H}_1}
  \right],\hspace{5mm}
  \label{eq:expansion}
\end{eqnarray}
where $\mathcal{H}_1= \mathcal{H}_\mathrm{loc} + \mathcal{H}_\mathrm{bath}$ and $\mathcal{H}_2= \mathcal{H}_\mathrm{mix}$ and we employed the interaction picture.

In Eq.~(\ref{eq:expansion}), the partition function $Z$ is represented as the sum of all configurations $c=\left\{ \tau_1, \cdots, \tau_n\right\}$ with 
weight 
\begin{eqnarray}
  && w_c=(-d\tau)^{n} \mathrm{Tr}
  \left[ 
  e^{-(\beta-\tau_n) \mathcal{H}_1}
  \mathcal{H}_2 
  e^{-(\tau_n-\tau_{n-1}) \mathcal{H}_1}
  \cdots
  \mathcal{H}_2 
  e^{-\tau_1 \mathcal{H}_1}
  \right]
  \mathrm{d} \tau^n.\hspace{5mm}
  \label{eq:w1}
\end{eqnarray}

The weight can be simplified further by exploiting the fact that the time evolution of the impurity and the bath are not coupled by $\mathcal{H}_2$.
By tracing out the bath degrees of freedom,
one obtains
\begin{eqnarray}
   w_{\tilde{c}} &=& Z_\mathrm{bath}  \mathrm{Tr_{loc}}\left[ e^{-\beta\mathcal{H}_\mathrm{loc}} T \opc_{\alpha_n}(\tau_n) \opcdag_{\alpha_n^\prime}(\tau_n^\prime) \cdots \opc_{\alpha_1}(\tau_1) \opcdag_{\alpha_1^\prime}(\tau_1^\prime)\right]\nonumber\\
  && \times \mathrm{det} \boldsymbol{M}^{-1}(\{\tau_1,\alpha_1\},\cdots,\{\tau_n,\alpha_n\};\{\tau_1^\prime,\alpha_1^\prime\},\cdots,\{\tau_n^\prime,\alpha_n^\prime\}) (\mathrm{d}\tau)^{2n}.\label{eq:w2}
\end{eqnarray}
Here, $\tilde{c}$ represents a configuration with annihilation operators at $\tau_1<\cdots<\tau_n$ with flavor $\alpha_1, \cdots, \alpha_n$ and creation operators at $\tau_1^\prime<\cdots<\tau_n^\prime$ with flavor $\alpha_1^\prime, \cdots, \alpha_n^\prime$.
The matrix element of $\boldsymbol{M}^{-1}$ at $(i,j)$ is given by the hybridization function $\Delta_{\alpha_i^\prime,\alpha_j}(\tau_i^\prime - \tau_j)$ defined in terms of $\epsilon_{k,\alpha}$ and $V_k^{\alpha,b}$.
The trace in Eq.~(\ref{eq:w2}) reduces to the form

\begin{eqnarray}
  && \mathrm{Tr_{loc}}
  \Big[ 
  e^{-(\beta-\tau_{2n}) \mathcal{H}_\mathrm{loc}}
  \hat{O}_{2n}
  e^{-(\tau_{2n}-\tau_{2n-1}) \mathcal{H}_\mathrm{loc}}
  \hat{O}_{2n-1}
  \cdots
  \hat{O}_1
  e^{-\tau_1 \mathcal{H}_\mathrm{loc}}
  \Big]=\nonumber\\
  &&\hspace{1cm} \sum_m \langle \Psi_m |
  e^{-(\beta-\tau_{2n}) \mathcal{H}_\mathrm{loc}}
  \hat{O}_{2n}
  e^{-(\tau_{2n}-\tau_{2n-1}) \mathcal{H}_\mathrm{loc}}
  \hat{O}_{2n-1}
  \cdots
  \hat{O}_1
  e^{-\tau_1 \mathcal{H}_\mathrm{loc}}
  |\Psi_m\rangle,\label{eq:trace}
\end{eqnarray}
where $\hat{O}_1,\cdots, \hat{O}_{2n}$ are time-ordered creation and annihilation operators appearing in Eq.~(\ref{eq:w2}).
$|\Psi_m\rangle$ denotes an eigenstate of $\mathcal{H}_\mathrm{loc}$, and the sum is over all eigenstates. 

The contributions of the configurations $\tilde{c}$ are stochastically sampled in the Monte Carlo simulation with the weight $w_{\tilde{c}}$. 
When $\mathcal{H}_\mathrm{loc}$ contains only chemical potentials and density-density interactions,
the occupation number basis is an eigensystem of $\mathcal{H}_\mathrm{loc}$.
In this case, Eq.~(\ref{eq:trace}) can be evaluated efficiently.
Otherwise, the evaluation of Eq.~(\ref{eq:trace}) is exponentially costly with respect to the number of orbitals in the impurity.

In Ref.~\cite{Lauchi09}, it was shown that the sum over eigen states can be restricted to ground states at low enough temperature.
It was also proposed to evaluate the trace using the so-called Krylov subspace method described in the next section.

\section{Imaginary time evolution with the Krylov subspace method}\label{sec:krylov}
In evaluating the trace in Eq.~(\ref{eq:trace}), we perform an imaginary time evolution
\begin{eqnarray}
  e^{-\tau \mathcal{H}} \boldsymbol{v}
\end{eqnarray}
in each time-interval between creation/annihilation operators.
We employ the Krylov subspace method in the same manner as in Ref.~\cite{Lauchi09}.

For a given Hamiltonian $\mathcal{H}$ and vector $\boldsymbol{v}$, the Krylov subspace is defined as
\begin{eqnarray}
  \mathcal{K}_p &=& \mathrm{span} \{ \boldsymbol{v}, \mathcal{H} \boldsymbol{v}, \cdots, \mathcal{H}^{p-1} \boldsymbol{v}\},
\end{eqnarray}
where $p$ is the dimension of the subspace.
Then, the full matrix exponential $e^{-\tau \mathcal{H}} \boldsymbol{v}$ is approximated by the matrix exponential of the Hamiltonian projected onto the Krylov space.

We construct an orthonormal basis for the Krylov subspace that tridiagonalizes $\mathcal{H}$ as
\begin{eqnarray}
  \boldsymbol{U}^\dagger \boldsymbol{H} \boldsymbol{U} &=& \boldsymbol{T} = 
  \left(
  \begin{array}{cccc}
        \alpha_1   &    \beta_1   &    0       & \cdots \\
        \beta_1    &    \alpha_2  & \beta_2    & \ddots \\
               0   &    \beta_2   &  \alpha_3  & \ddots \\
        \vdots   &      \ddots   & \ddots    & \ddots \\
      \end{array}
  \right)
\end{eqnarray}
by using the Lanczos method.
Here, $\alpha_i$ and $\beta_i$ are real numbers. The 
column vectors of $\boldsymbol{U}$ are orthonormal basis vectors $\{ \boldsymbol{u}_i\}$ with $\boldsymbol{u}_1=\boldsymbol{v}/\| \boldsymbol{v}\|$.

The basis vectors $\boldsymbol{u}_i$ and the matrix elements $\alpha_i$, $\beta_i$ are obtained step by step for $i=1,2,3,\cdots$ as follows:
\begin{eqnarray}
  \alpha_i &=& \boldsymbol{u}_i^\dagger \boldsymbol{H} \boldsymbol{u}_i, \label{eq:lanczos-step1} \\
  \boldsymbol{v}_{i+1} &=& \left\{
       \begin{array}{ll}
       \boldsymbol{H} \boldsymbol{u}_i - \alpha_i \boldsymbol{u}_i & (i=1)\\
       \boldsymbol{H} \boldsymbol{u}_i - \beta_{i-1} \boldsymbol{u}_{i-1} - \alpha_i \boldsymbol{u}_i & (i>1) \\
       \end{array}
       \right.,\label{eq:lanczos-step2} \\
  \beta_i &=& ||\boldsymbol{v}_{i+1}||, \\
  \boldsymbol{u}_{i+1} &=& \boldsymbol{v}_{i+1}/\beta_i.
\end{eqnarray}

Convergence of the result is checked at each Lanczos step between Eqs.~(\ref{eq:lanczos-step1}) and (\ref{eq:lanczos-step2}) by evaluating the matrix exponential as 
\begin{eqnarray}
  e^{-\tau \boldsymbol{H}} \boldsymbol{v} &=& \beta_0 e^{-\tau \boldsymbol{H}} \boldsymbol{u}_1= \beta_0 \sum_{i=1}^{p} \left( e^{-\tau \boldsymbol{T}} \right)_{i1} \boldsymbol{u}_i,
\end{eqnarray}
where $\beta_0 = \| \boldsymbol{v} \|$.
The matrix exponential $e^{-\tau \boldsymbol{T}}$ can be evaluated by a direct diagonalization because of the small dimension of the Krylov subspace.
In the following calculations, we use the criterion $|\left( e^{-\tau \boldsymbol{T}} \right)_{m1}/\left( e^{-\tau \boldsymbol{T}} \right)_{11}|<\epsilon$ with the torelance $\epsilon=10^{-5}$.

\section{Quantum impurity model}\label{sec:model}
Throughout this paper, we consider an $N$-orbital impurity model with a ``Slater-Kanamori" interaction. 
The Hamiltonian is
\begin{eqnarray}
  \mathcal{H}_\mathrm{loc} &=& \sum_i U \hat{n}_{i\uparrow} \hat{n}_{i\downarrow} - \mu \sum_i \hat{n}_i \nonumber \\
  && + \sum_{i>j, \sigma} \left[U^\prime \hat{n}_{i\sigma} \hat{n}_{j-\sigma} + (U^\prime-J) \hat{n}_{i\sigma} \hat{n}_{j\sigma}\right]\nonumber \\
  && - \sum_{i \neq j} J \left( \hat{c}^\dagger_{i\downarrow}  \hat{c}^\dagger_{j\uparrow} \hat{c}_{j\downarrow} \hat{c}_{i\uparrow} + \hat{c}^\dagger_{j\uparrow}  \hat{c}^\dagger_{j\downarrow} \hat{c}_{i\uparrow} \hat{c}_{i\downarrow}\right),\label{eq:imp}
\end{eqnarray}
where $\hat{c}^\dagger_i$ and $\hat{c}_i$ are creation/annihilation operators of an electron at site $i$, and $\hat{n}_i\equiv \hat{c}^\dagger_i\hat{c}_i$.
We take $U^\prime = U-2J$ and $J=U/6$.
The chemical potential is chosen such that the system is at half filling: $\mu = (n-\frac{1}{2})U-(n-1)\frac{5}{2}J$.
We consider an orbital-diagonal hybridization function corresponding to a noninteracting model with semicircular density of states of bandwidth 4.

While the interaction terms in Eq.~(\ref{eq:imp}) may not correspond to a rotationally invariant interaction for $N>3$, we use this Hamiltonian for the purpose of benchmark calculations. 
We do not take into account the special conserved quantities~\cite{Parragh012} which enable a particularly efficient sampling of the Slater-Kanamori Hamiltonian.
None of the procedures discussed in the following sections depend on a specific form of the Hamiltonian.

\section{Trace calculation with matrix product states}\label{sec:krylov-mps}
In this section, we investigate the accuracy and efficiency of a combined Krylov and MPS approach. 
A brief introduction of the MPS formalism is given in Sec.~\ref{sec:mps}.
In Sec.~\ref{sec:numerics}, we describe the details of benchmark calculations.
In Sec.~\ref{sec:mps-accuracy} we discuss the accuracy of the method, while 
the performance of the method is investigated in Sec.~\ref{sec:mps-performance}.
Future perspectives are given in Sec.~\ref{sec:mps-discussion}.

\subsection{Matrix product state formalism}\label{sec:mps}
Here we provide a very brief overview of the MPS formalism.
For details see the review by Schollw\"{o}ck.~\cite{Schollwock-review}
\subsubsection{Matrix product states (MPS)}
Let us consider a one-dimensional lattice of length $L$,
with a local Hilbert space of dimension $d$ at each site.
Hereafter, the dimension $d$ is referred to as the local dimension.
For instance, Hubbard models with $S=1/2$ electrons have a local dimension $d=4$: The local Hilbert space at site $i$ can be spanned by 
$|0\rangle$, $\opcdag_{i\downarrow} |0\rangle$, $\opcdag_{i\uparrow} |0\rangle$, $\opcdag_{i\uparrow} \opcdag_{i\downarrow} |0\rangle$.

Any pure state can be represented in the form
\begin{eqnarray}
  |\Psi\rangle \nonumber &=& \sum_{\sigma_1,\cdots,\sigma_L} c_{\sigma_1,\cdots,\sigma_L} |\sigma_1,\cdots,\sigma_L \rangle \nonumber\\
  &=& \sum_{\sigma_1,\cdots\sigma_L} \left(\sum_{b_1,\cdots,b_L}^{r_1,\cdots,r_L} M^{\sigma_1}_{1,b_1} M^{\sigma_2}_{b_1,b_2} \cdots M^{\sigma_L}_{b_{L-1},b_L}\right)\nonumber \\
  && \times |\sigma_1\cdots \sigma_L\rangle\nonumber\\
  &=& \sum_{\sigma_1,\cdots,\sigma_L}\boldsymbol{M}^{\sigma_1} \boldsymbol{M}^{\sigma_2} \cdots \boldsymbol{M}^{\sigma_L} |\sigma_1\cdots \sigma_L\rangle,\label{eq:mps}
\end{eqnarray}
where $\boldsymbol{M}^{\sigma_l}$ ($l=1,\cdots,L$) are rank-3 tensors of dimension $d\times r_{l-1} \times r_{l}$.
At the left ($l=1$) and right ($l=L$) edges, we take $r_0=r_{L+1}=1$.
The maximum value of $b_l$ is referred to as the bond dimension of the MPS.

The MPS formalism is the underlying variational approximation made by the DMRG algorithm.~\cite{Schollwock-review}
For a non-critical 1D system with short-range interactions,
the ground state can be described very accurately by an MPS with a small bond dimension of $O(1)$.
Note that the exponentially large tensor $c_{\sigma_1,\cdots,\sigma_L}$ is reduced to a product of small tensors of size $O(1)$ because the entanglement entropy of the ground state is $O(1)$ with respect to the system length. 

\subsubsection{Compressing MPS}
An important remark is that MPS with a fixed bond dimension do not form a vector space.
For example, the sum of two MPS results in a larger bond dimension as discussed later in Sec.~\ref{sec:algebra}.
In general, an MPS with a larger bond dimension can contain more information.
Thus, to keep the bond dimension bounded, one may have to reduce the bond dimension after an operation, while keeping the loss of accuracy as small as possible.
This can be done by an algorithm based on the so-called singular value decomposition (SVD).
A truncation of the bond dimension from $D^\prime$ to $D$ costs $O(dD^{\prime 3}L)$ for $D^\prime \gg D$.

\subsubsection{Matrix product operators (MPO)}
Matrix product operators are a natural generalization of the MPS concept to operators.
Let us consider an arbitrary operator $\hat{O}$:
\begin{eqnarray}
  \hat{O} &=& \sum_{\boldsymbol{\sigma},\boldsymbol{\sigma}^\prime} O_{\bsigma,\bsigma^\prime} |\boldsymbol{\sigma}\rangle \langle \boldsymbol{\sigma}^\prime|.
\end{eqnarray}
The idea of MPS is directly applicable to operators by regarding $(\sigma_l\sigma_l^\prime)$ as one big index at each site.
That is, the coefficients are represented as a product of local tensors as follows:
\begin{eqnarray}
  O_{\bsigma,\bsigma^\prime}  &=& \sum_{b_1,\cdots,b_L}^{r_1,\cdots,r_L} {W}^{\sigma_1\sigma_1^\prime}_{1,b_1} {W}^{\sigma_2\sigma_2^\prime}_{b_1,b_2} \cdots {W}^{\sigma_L\sigma_L^\prime}_{b_{L-1},b_L} \nonumber\\
    &=& \boldsymbol{W}^{\sigma_1\sigma_1^\prime} \boldsymbol{W}^{\sigma_2\sigma_2^\prime} \cdots \boldsymbol{W}^{\sigma_L\sigma_L^\prime},
\end{eqnarray}
where the $\bW$'s are now rank-4 tensors.
The maximum value of $b_l$ is referred to as the bond dimension of the MPO.
We discuss how to construct an MPO for a given Hamiltonian in Sec.~\ref{sec:numerics}.

\subsubsection{Linear algebra with MPS and MPO}\label{sec:algebra}
We can perform fundamental operations in quantum mechanics in the framework of MPS and MPO.
One of the simplest examples is the summation of two wavefunctions $|\phi_1\rangle$ and $|\phi_2\rangle$,
as is required in Eqs.~(\ref{eq:lanczos-step1}) and (\ref{eq:lanczos-step2}).
The sum of two MPSs with bond dimensions $D_1$ and $D_2$, respectively, has a bond dimension of $D^\prime \le D_1 + D_2$.
This can be understood by considering the sum of two MPSs with bond dimension one:
\begin{eqnarray}
  |\phi_1\rangle &=& \sum_{\bsigma} A^{\sigma_1} \cdots A^{\sigma_L} |\bsigma\rangle, \\
  |\phi_2\rangle &=& \sum_{\bsigma} B^{\sigma_1} \cdots B^{\sigma_L} |\bsigma\rangle.
\end{eqnarray}
One can easily see that the sum is given by
\begin{eqnarray}
  && \sum_{\bsigma} (A^{\sigma_1} B^{\sigma_1})
    \left(\begin{array}{cc} A^{\sigma_2} &0 \\ 0 & B^{\sigma_2} \end{array}\right)
      \times \cdots
      \times\left(\begin{array}{cc} A^{\sigma_{L-1}} &0 \\ 0 & B^{\sigma_{L-1}} \end{array}\right) 
    \left(\begin{array}{c} A^{\sigma_L} \\ B^{\sigma_L} \end{array}\right)
     |\bsigma\rangle,\nonumber\\
\end{eqnarray}
with a bond dimension of two.
This can be extended to larger bond dimensions in a straightforward way.
A sum of two MPS of bond dimension $D$ requires only $O(dD^2L)$ operations.
However, it may be necessary to compress the resulting MPS to keep the bond dimension bounded at $D$.
This cost dominates over the summation for $D\gg 1$ because the compression is $O(dD^3L)$.

Another important operation is applying an operator $\hat{O}$ to a wavefunction $|\phi\rangle$, such as applying the Hamiltonian to a wavefunction in Eq.~(\ref{eq:lanczos-step2}).
Let us consider an MPO of bond dimension $D_W$ and an MPS of bond dimension $D$.
In this paper, we adopt an iterative approach which minimizes the residual $\| |\tilde{\phi}\rangle - \hat{O} |  \phi\rangle \|^2$ with respect to $|\tilde{\phi} \rangle$ for a fixed bond dimension $D$.
This algorithm scales as $O(LD^3D_Wd)$ for $1 \ll D_W \ll D$.~\cite{Schollwock-review}

\subsection{Numerical details}\label{sec:numerics}
The simulations in this section are carried out for the impurity model given in Sec.~\ref{sec:model}.
We take $U=6$ and $J=U/6$ and $\beta=50$.
The Hamiltonian (\ref{eq:imp}) can be represented by an MPO with a bond dimension of $D_\mathrm{W}\propto \Norb^2$ 
because the MPO for each term in Eq.~(\ref{eq:imp}) has a bond dimension of 1.
This means that the computational effort scales polynomially with $\Norb$ as $O(D^3\Norb^3)$.
A further speed-up can be achieved by compressing the MPO.
As will be explained in 
\ref{sec:mpo-compression},
the Hamiltonian can be represented by a more compact MPO with bond dimension eight irrespective of $\Norb$ because the two-body interactions are homogeneous.
Using this compact MPO, the computational effort now scales as $O(D^3\Norb)$.

Although we consider the Slater-Kanamori interaction in this paper,
the approach can be applied to any impurity model including general one- and two-body interactions like intra-orbital hopping and correlated hopping.
As explained in Appendix~\ref{sec:mpo-general-int}, any one- and two-body interaction term can be represented by an MPO with bond dimension one.
The compression of the MPO for the Hamiltonian is also possible for general one- and two-body interactions.

The following calculations were performed on a 2GHz Intel Core i7 CPU (Ivy Bridge), without parallelization.
We used the Intel C++ Compiler v13.0 and the Math Kernel Library.
The imaginary time evolution was implemented with the MAQUIS/DMRG code.~\cite{MAQUIS}
The following results were obtained without exploiting good quantum numbers such as the total electron number.
We found that exploiting conserved quantum numbers does not reduce the computational cost for the small bond dimensions $D\le 50$ used in this study.

We measured the timings and the accuracy using the Krylov-sparse-matrix and Krylov-MPS methods as follows.
First, we perform Monte Carlo simulations with an exact solver (Krylov-sparse-matrix solver) in the same way as in Ref.~\cite{Lauchi09}.
After thermalization, we randomly select several configurations and measure timings.
Calculations with the MPS method are then repeated for the same configurations using the MAQUIS/DMRG code.
In the following, we measure the timings and the accuracy of the imaginary-time evolution for the ground state of the largest subspace with $(N_\uparrow,N_\downarrow)=(3,2)$, (4,3), (4,4), (5,5) for $\Norb$=5, 6, 7, 8, and 10, respectively.

\begin{figure}[h]
 \centering
 \includegraphics[width=.5\textwidth,clip]{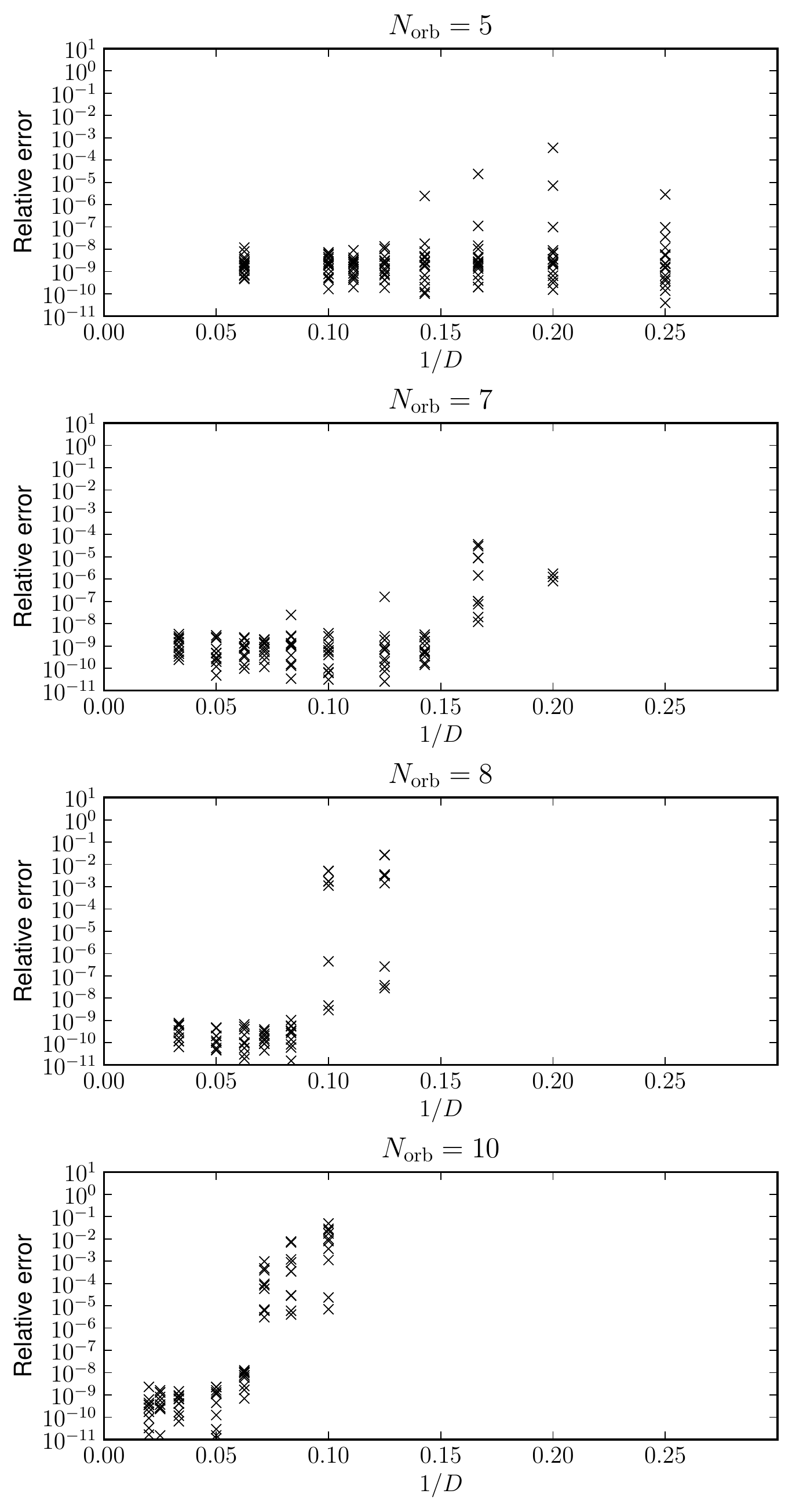}
 \caption{Convergence of the value of the trace with respect to the bond dimension $D$ for $N_\mathrm{orb}=5$, 7, 8, 10, respectively.}
 \label{fig:1}
\end{figure}

\subsection{Accuracy of the MPS method}\label{sec:mps-accuracy}
First, we discuss the accuracy of the MPS formalism.
In Fig.~\ref{fig:1}, we show the convergence of the calculated value of the trace with respect to the bond dimension $D$.
The impurity sizes are $\Norb=5$, 7, 8, and 10.
The relative error is defined as $|(t(D)-t_\mathrm{exact})/t_\mathrm{exact}|$,
where $t(D)$ and $t_\mathrm{exact}$ are the values of the trace calculated by the MPS formalism and the exact solver, respectively.
The expansion order per flavor $N_\mathrm{exp}$ is 4.5--5 for $\Norb=5$ and 3--3.5 for $\Norb=$ 7, 8, and 10, respectively.

As seen in Fig.~\ref{fig:1}, the relative error of the MPS method decreases rapidly as $D$ increases.
For $\Norb=5$, the results are already converged at $D=8$ for all sets of $\{\hat{O}(\tau_i)\}$. For the largest system, i.e., $\Norb=10$, the relative error is well converged (and below $10^{-7}$) at $D=16$, even though the dimension of the Hilbert space is $(_{10}\mathrm{C}_5)^2 = 63,504$.
These results show that the MPS formalism yields accurate results even with a bond dimension considerably smaller than the dimension of the Hilbert space.

\subsection{Performance of the MPS method}\label{sec:mps-performance}
\begin{figure}[h]
 \centering
 \includegraphics[width=.495\textwidth,clip]{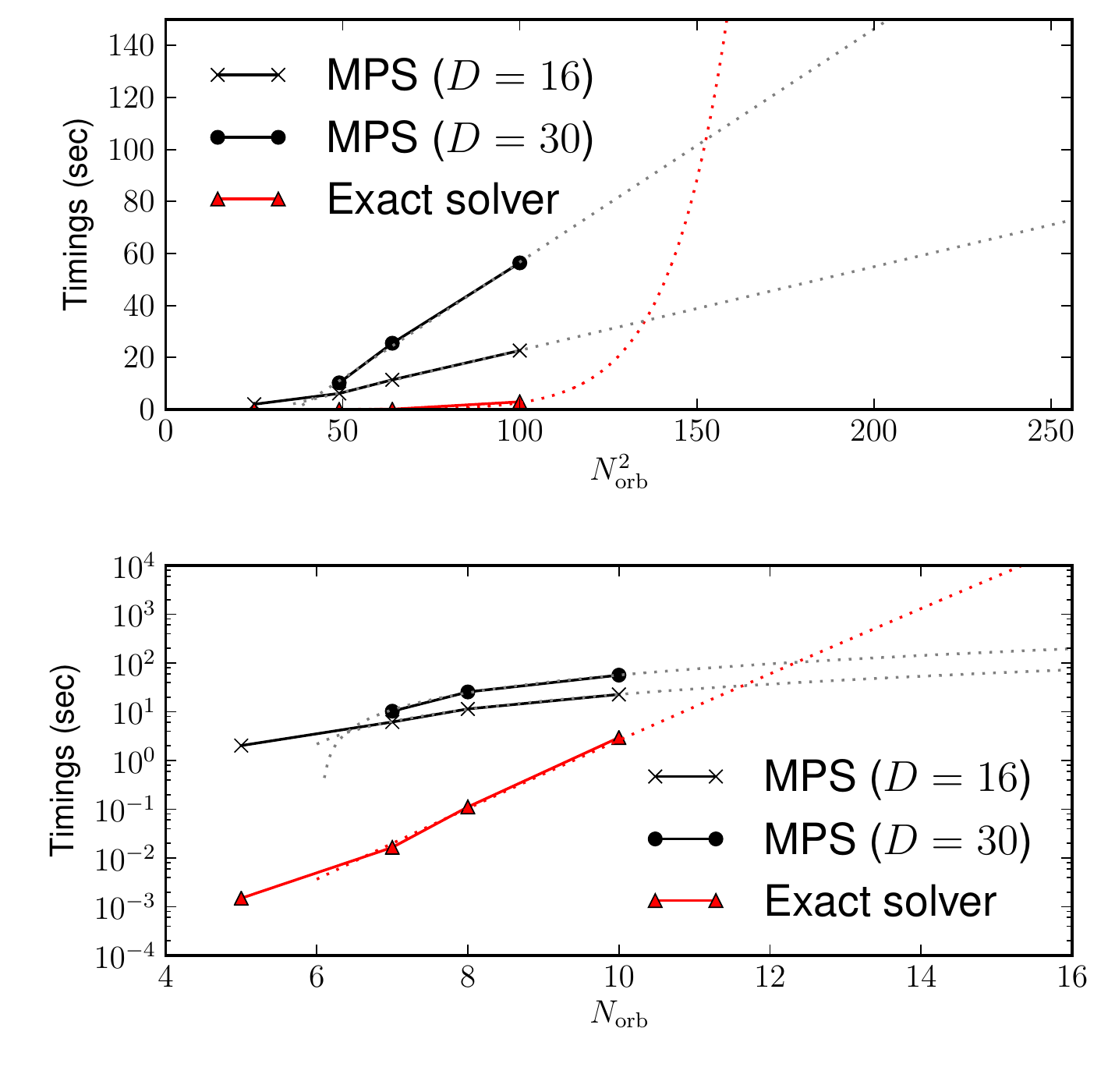}
 \caption{$\Norb$ dependence of the timings for the imaginary time evolution in the interval $[0, \beta]$.
 The data are averaged over 10 different operator configurations $\{\hat{O}(\tau_i)\}$ for each $\Norb$.
 The red broken line and the dotted black line are the fit by Eqs.~(\ref{eq:exp}) and (\ref{eq:mps-scaling}), respectively.
 }
 \label{fig:timing}
\end{figure}

Next, we compare the performance of the two methods.
Figure~\ref{fig:timing} shows the timing for an imaginary time evolution in the interval $[0,\beta]$.
It is clearly seen that the timing for the exact solver increases exponentially with $\Norb$.
The red broken line in Fig.~\ref{fig:timing} is a fit by
\begin{eqnarray}
  C \Norb^2 4^\Norb,\label{eq:exp}
\end{eqnarray}
where $C$ is a positive constant.
Equation~(\ref{eq:exp}) is derived 
as follows.
The most costly operation in the imaginary time evolution is applying a sparse matrix $\mathcal{H}$ to a dense vector in Eqs.~(\ref{eq:lanczos-step1}) and (\ref{eq:lanczos-step2}).
Each such operation costs $O(\Norb^2 D_\mathrm{Hilbert})$, where the dimension of the largest subspace $D_\mathrm{Hilbert}$ is given by $(_\Norb\mathrm{C}_{\Norb/2})^2\propto 4^\Norb/\Norb$.
Assuming that the expansion order per orbital is $O(1)$, we immediately arrive at Eq.~(\ref{eq:exp}).
As shown in Fig.~\ref{fig:timing}, the data are well fitted by Eq.~(\ref{eq:exp}) for $\Norb\ge 7$ with $C=2.5\times 10^{-8}$.

\begin{figure}[h]
 \centering
 \includegraphics[width=.45\textwidth,clip]{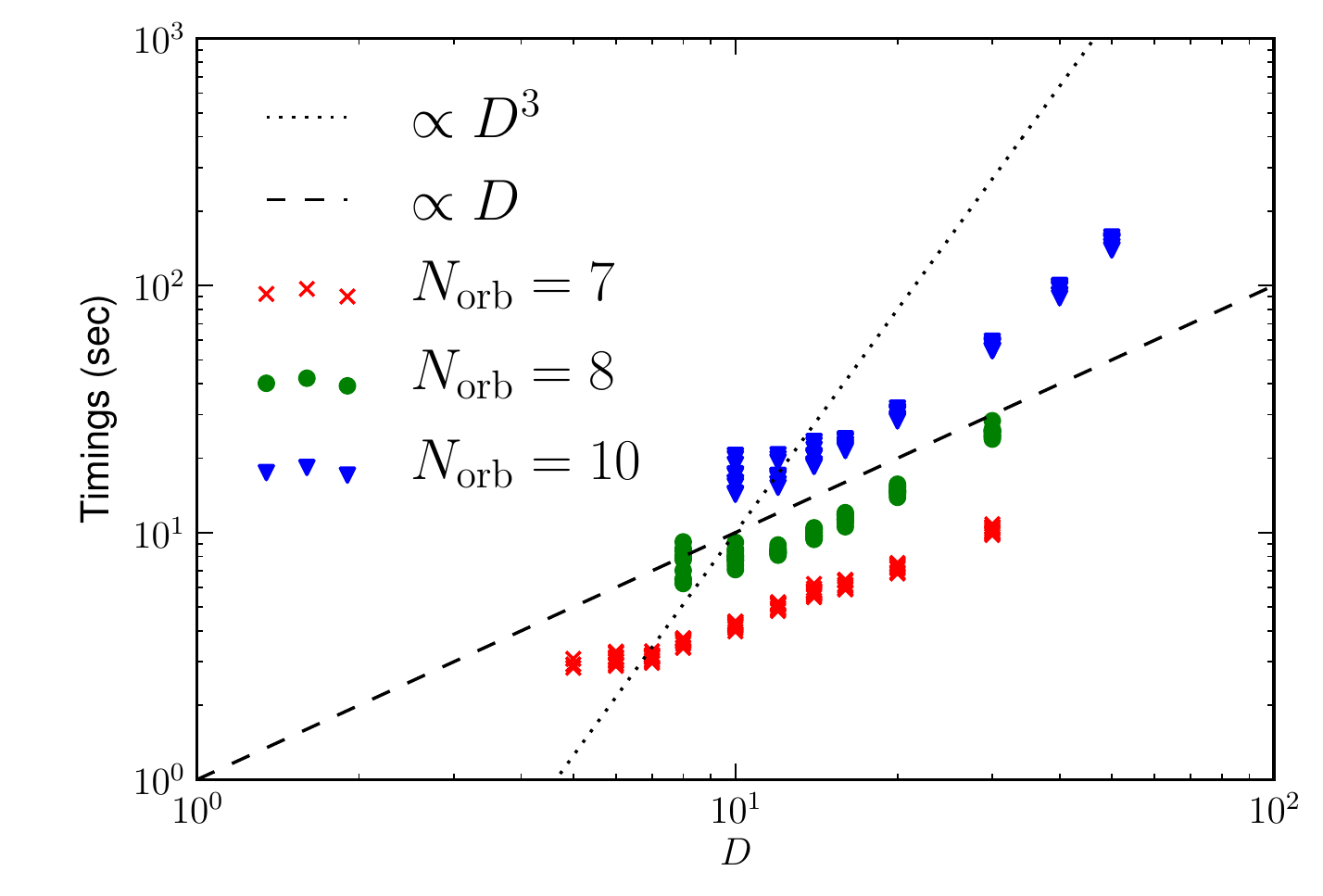}
 \caption{Bond-dimension $D$ dependence of the timings for an imaginary time evolution in the interval $[0, \beta]$.
 Data for $\Norb=7$, 8, and $10$ are shown.
 The different points represent data for different operator configurations $\{\hat{O}(\tau_i)\}$.}
 \label{fig:timing-D}
\end{figure}

On the other hand, the timing for the MPS formalism is expected to scale as $O(D^3 \Norb^2)$ for a fixed $D$.
This comes from the fact that applying $\mathcal{H}$ to an MPS costs $O(D^3 \Norb)$.
As seen in Fig.~\ref{fig:timing}, the data are indeed well fitted by the expected scaling
\begin{eqnarray}
  a (\Norb^2 - b)\label{eq:mps-scaling}
\end{eqnarray}
with $a$ and $b$ positive constants.
We note that the estimated value of $a$ increases only slightly from 0.322 to 0.896 as $D$ increases from $D=16$ to $D=30$,
though one expects a $(30/16)^3~(\simeq 6.59)$ time increase. 
This may be due to overhead in treating many small matrices for small $D$.
This can be seen more explicitly when we plot the timings as a function of $D$ for each $\Norb$ in Fig.~\ref{fig:timing-D}.
It is obvious that the timing increases more slowly than the expected asymptotic scaling $O(D^3)$ for $D\le 50$.

Even for the largest $\Norb$ considered ($\Norb=10$), the MPS formalism with $D=16$ runs about 10 times slower than the exact solver.
However, the MPS formalism is expected to become more efficient than the exact solver for larger $\Norb$.
Extrapolating the timings of the two methods using Eqs.~(\ref{eq:exp}) and (\ref{eq:mps-scaling}),
the crossover point is estimated to be $\Norb=12$--$13$,
with only a slight dependence on the value of $D$ (see the lower panel of Fig.~\ref{fig:timing-D}).

\subsection{Discussion and future perspectives}\label{sec:mps-discussion}
Our results show that the Krylov-MPS formalism can be potentially superior to the exact Krylov-sparse-matrix solver for large number of orbitals $\Norb\gtrsim 12$.
Impurity problems with $\Norb \ge 12$ are relevant for example for cluster-type DMFT calculations of multi-orbital Hubbard models. 
However, the MC simulation of such large impurity problems is not feasible at the moment with our present code.
(To date, simulations with hybridization expansion solvers have been restricted to at most 7 orbitals.)
Thus, in this section, we discuss how the performance might be improved.

In a MC simulation,
we update $\{\hat{O}(\tau_i)\}$ by an elementary update such as inserting or removing a pair of annihilation and creation operators.
Each operator must be updated before the MC sampling loses its memory of the original configuration. 
Thus, the autocorrelation time $\tau_\mathrm{auto}$ is expected to be roughly $2 \Norb N_\mathrm{exp}/p_\mathrm{acc}$ in units of elementary updates ($N_\mathrm{exp}$ is the expansion order per flavor).
The acceptance rate $p_\mathrm{acc}$ depends on the system and on parameters such as $\beta$. It is typically on the order of 0.01--0.1.
Assuming $p_\mathrm{acc}=0.1$ and $N_\mathrm{exp}=3$ (a typical value in the strongly correlated regime, and for temperatures of about 1\% of the bandwidth) for $\Norb=12$, we obtain $\tau_\mathrm{auto} \simeq 700$ elementary updates.
Recalling that the timing for evaluating the trace is $O(10^2)$ seconds (see Fig.~\ref{fig:timing}),
the autocorrelation time is estimated to be $O(10^5)$ seconds or 30 hours.
In a weakly correlated metal, where the perturbation order is higher, the autocorrelation time is on the order of a week.
This is too long for practical DMFT calculations.

There are possible ways to reduce the autocorrelation time.
First, we can increase the acceptance rate $p_\mathrm{acc}$ by proposing several candidates at each MC update.
Evaluating their weights can be assigned to different nodes.
By using the heat bath algorithm or a better algorithm,~\cite{Suwa-Todo-PRL} the acceptance rate $p_\mathrm{acc}$ can be increased to almost 1.
Another factor of 10 can be gained by using the improved MC sampling introduced in Sec.~\ref{sec:improved-sampling},
which avoids recomputing the full imaginary-time evolution from scratch.
By using these two tricks, the autocorrelation time $\tau_\mathrm{auto}$ can be redued to $O(10^3)$ seconds,
which is still not short enough for practical applications.

Another possible way is to speed up the imaginary time evolution by parallel computing.
However, this is not trivial because the bond dimensions of MPS and MPO tensors are quite small in the case of quantum impurity problems.

\section{Improved Monte Carlo sampling}\label{sec:improved-sampling}
In this section, we propose an improved  Monte Carlo sampling procedure which significantly reduces autocorrelation times for multi-orbital impurity problems.
This sampling strategy can be used both in the Krylov-sparse-matrix method and the Krylov-MPS method.
After reviewing previously proposed improved sampling strategies in Sec.~\ref{sec:smpl-conventional-methods}, we describe our new MC sampling scheme in Sec.~\ref{sec:exp-window}.
We explain the details of benchmark calculations in Sec.~\ref{sec:smpl-setup}.
The benchmark results are shown in Sec.~\ref{sec:smpl-result}, 
and future perspectives are discussed in Sec.~\ref{sec:smpl-discussion}. 
\begin{figure}[h]
 \centering
 \includegraphics[width=.35\textwidth,clip]{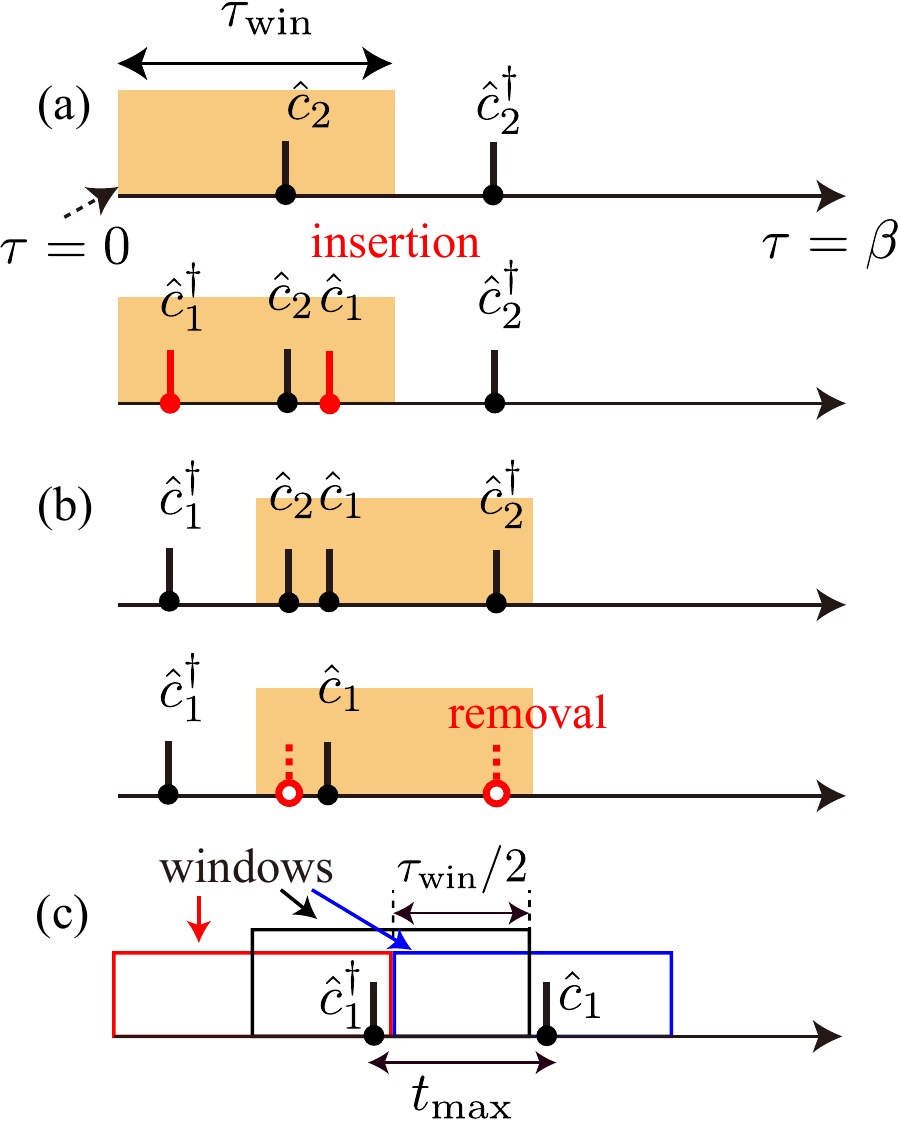}
 \caption{
 (a) Inserting a pair of operators in the window. Updates are allowed only in the window.
 (b) The window is moved by $\tau_\mathrm{win}/2$ from (a). Now, the pair $\{\hat{c}_2^\dagger,\hat{c}_2\}$ can be removed.
 (c) The pair does not fit into any of the three windows. This can happen if $\tau_\mathrm{win}<2t_\mathrm{max}$. 
 }
 \label{fig:localwindow}
\end{figure}

\subsection{Conventional method}\label{sec:smpl-conventional-methods}
In a hybridization-expansion continuous-time Monte Carlo simulation,
one updates a current configuration by proposing a new configuration which is slightly different from the current one:
For example, one tries to insert or remove a pair of creation and annihilation operators.
Then, the new configuration is accepted stochastically according to the ratio of the weights of the current and new configurations.
Naively, one might evaluate the weight of the new configuration by performing an imaginary time evolution in the full interval $[0,\beta]$.
However, the computational cost of such a calculation grows linearly with the expansion order $N_\mathrm{exp}$.
This is costly at low temperature or in a metallic phase, where the expansion order is large. 

For the matrix formalism, in which the trace is evaluated using matrix products,
Haule proposed a trick to improve the efficiency of the Monte Carlo sampling.~\cite{Haule07}
Here, we introduce a related idea in the context of the Krylov method.
The improved sampling strategy proposed in Ref.~\cite{Haule07} is based on the observation that the insertion or removal of pairs of operators is predominantly a local (in imaginary time) process. 
In other words, the acceptance rate for an insertion or removal of a pair of operators with a large time difference is very low.
It was furthermore proposed to do the time evolution of wave vectors from both sides, storing the resultant vectors (or matrix products) at several intermediate $\tau$ points.
Then, when trying to insert or remove a pair of operators with a short time difference,
the evaluation of the trace only requires the time evolution in a short time interval.
Although this makes trial steps cheaper, one has to recompute the intermediate results once an update is accepted.
Since this requires the time evolution from both sides, which typically costs $O(N_\mathrm{exp})$, the method does not change the scaling with respect to $N_\mathrm{exp}$.
Thus, the method gives a significant improvement in performance only when the acceptance rate is quite low and $N_\mathrm{exp}$ is not so large.

\subsection{Sliding-window approach}\label{sec:exp-window}
We now propose an improved update scheme in which the computational cost of an elementary update stays constant with respect to 
the expansion order $N_\mathrm{exp}$.
Although the exponential scaling with the number of orbitals is not affected, this method substantially reduces the prefactor of the scaling at low temperatures.
Although we explain the idea in the context of the Krylov algorithm, it can be applied to the matrix formalism as well.

First, to make the maximum use of the locality in the imaginary time,
we introduce an upper bound $t_\mathrm{max}$ on the time difference between the two operators which we try to insert or remove.
As mentioned in the previous study, $t_\mathrm{max}$ can be almost independent of $\beta$ and $N_\mathrm{exp}$.
In addition,
we introduce an imaginary-time window in which updates are allowed [see Fig.~\ref{fig:localwindow}(a)].
The window width $\tau_\mathrm{win}=\beta/N_\mathrm{win}$ is taken to be larger than (but on the order of) $t_\mathrm{max}$.
Now, similarly to Ref.~\cite{Haule07}, one performs the time evolution from both sides and stores the results at the end-points of the window.
This allows us to evaluate the trace for a new configuration at constant cost.

After several updates, the window is moved to the next position by $\tau_\mathrm{win}/2$ [see Fig.~\ref{fig:localwindow}(b)].
Concurrently, one updates the wave vectors at the end-points, which again costs only $O(\tau_\mathrm{win})=O(1)$.
This procedure is repeated so that the window moves back and forth in the whole interval $[0,\beta]$. 
This procedure is ergodic because we can produce operator pairs with arbitrary separation by inserting one with a short separation and then gradually increasing the separation through MC updates. 
We refer to \ref{sec:ergodicity} for a proof of ergodicity.
The advantage of our algorithm is that the cost of each MC step is independent of $\beta$ and 
reperforming the time evolution over the full time interval is not required.

By definition, $\tau_\mathrm{win}$ needs to be larger than $t_\mathrm{max}$.
In practical simulations, however, $\tau_\mathrm{win}$ has a stricter lower bound:
\begin{eqnarray}
  \tau_\mathrm{win}>2t_\mathrm{max}.\label{eq:lower-bound}
\end{eqnarray}
Let us consider a pair of operators with a time difference of $t_\mathrm{max}$ as shown in Fig.~\ref{fig:localwindow}(c).
When $\tau_\mathrm{win}<2t_\mathrm{max}$, this pair does not fit into any of the local windows shown there, 
and thus cannot be removed by a single elementary update.
Although this does not break the ergodicity, the autocorrelation time may increase. 
This problem can be avoided by taking $\tau_\mathrm{win}>2t_\mathrm{max}$.

If the window moves from one side to the other fast enough compared to the autocorrelation time,
the sequential sweep should not badly affect the autocorrelation time.
As discussed in Sec.~\ref{sec:mps-discussion}, the autocorrelation time is O($N_\mathrm{exp}N_\mathrm{orb}/p_\mathrm{acc}$).
On the other hand, moving the window from one side to the other takes $2N_\mathrm{win}=O(\Norb N_\mathrm{exp})$ MC steps.
Because these two time scales are always of the same order, the autocorrelation time should not be severely affected.

\subsection{Benchmark setup}\label{sec:smpl-setup}
\begin{figure}[!]
 \centering
 \includegraphics[height=.8\textheight,clip]{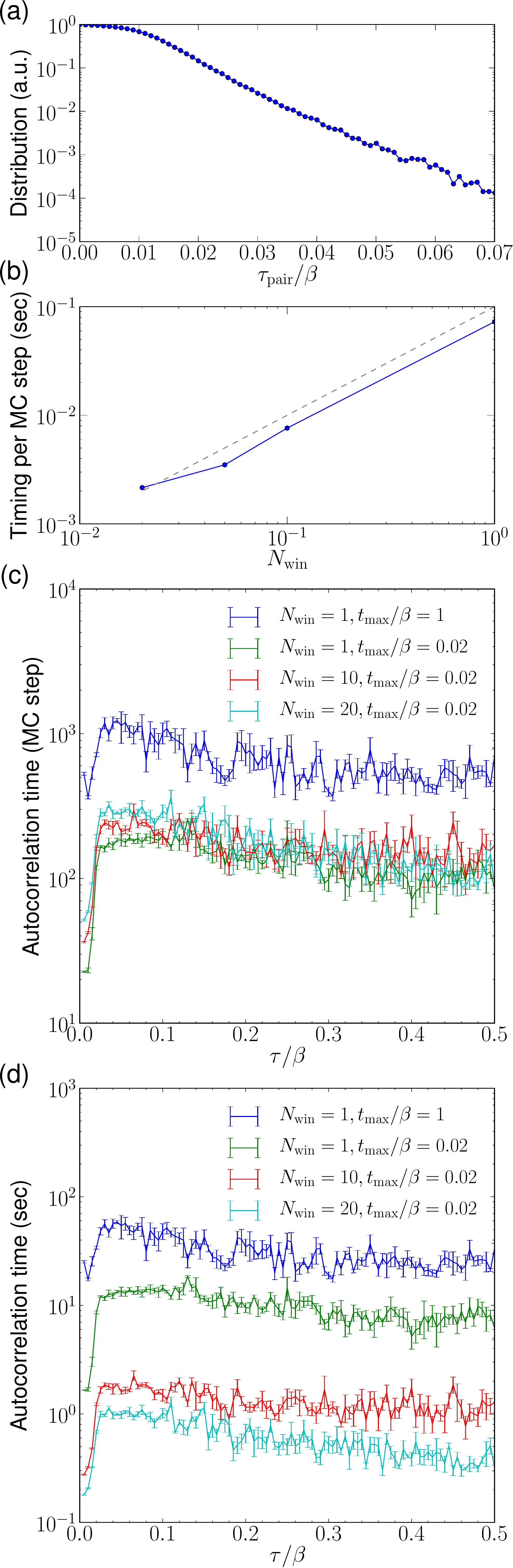}
 \caption{
 Benchmark results for a 5-orbital model with $U=6$ and $J=1$.
 (a) Distribution of the time difference of a successfully removed or inserted pair of operators in the MC sampling performed with $=t_\mathrm{max}/\beta=1$ and $N_\mathrm{win}=1$.
 (b) Timing per MC step. The broken line represents the relation  $\mathrm{timing} \propto \tau_\mathrm{win}$. 
 (c)/(d) Autocorrelation time of $G(\tau)$ in units of MC steps [(c)] and seconds [(d)].
 }
 \label{fig:autocorr}
\end{figure}
In this section, we show benchmark results for a 5-orbital impurity problem. 
At each position of the window, we try to insert or remove a pair $N_\mathrm{f}$ times, where $N_\mathrm{f}$ is the number of flavors. 
When trying to insert a pair, the time difference between the operators is chosen randomly in the interval $[0,t_\mathrm{max}]$.
Correspondingly, when we try to remove a pair in the window, we first list all pairs of creation and annihilation operators with a time difference equal to or less than $t_\mathrm{max}$.
Then, we try to remove one of them.
The detailed procedure is described in \ref{sec:MC-detail}.

The following simulations were done by the Krylov algorithm based on sparse-matrix techniques on one CPU core of AMD Opteron 6174 (2.2 GHz).
We divide the Hilbert space into sectors by using the total particle number $\hat{N}$ and magnetization $\hat{S}_z$ as good quantum numbers. 
All data are averaged over 4 independent MC runs of fixed $1.28\times 10^7$ steps.
The diagonal Green's function $G(\tau)$ is measured on 1001 points on the imaginary-time interval.
We symmetrize $G(\tau)$ by using the particle-hole symmetry.
The autocorrelation time of $G(\tau)$ is estimated by a binning analysis for each time point using bins of 16384 MC steps.
Simulations were performed using the ALPS libraries.~\cite{alps2}

\subsection{Benchmark results}\label{sec:smpl-result}
\subsubsection{Insulating region: $U=6$ and $\beta=50$}
\begin{figure}[!]
 \centering
 \includegraphics[width=.4\textwidth,clip]{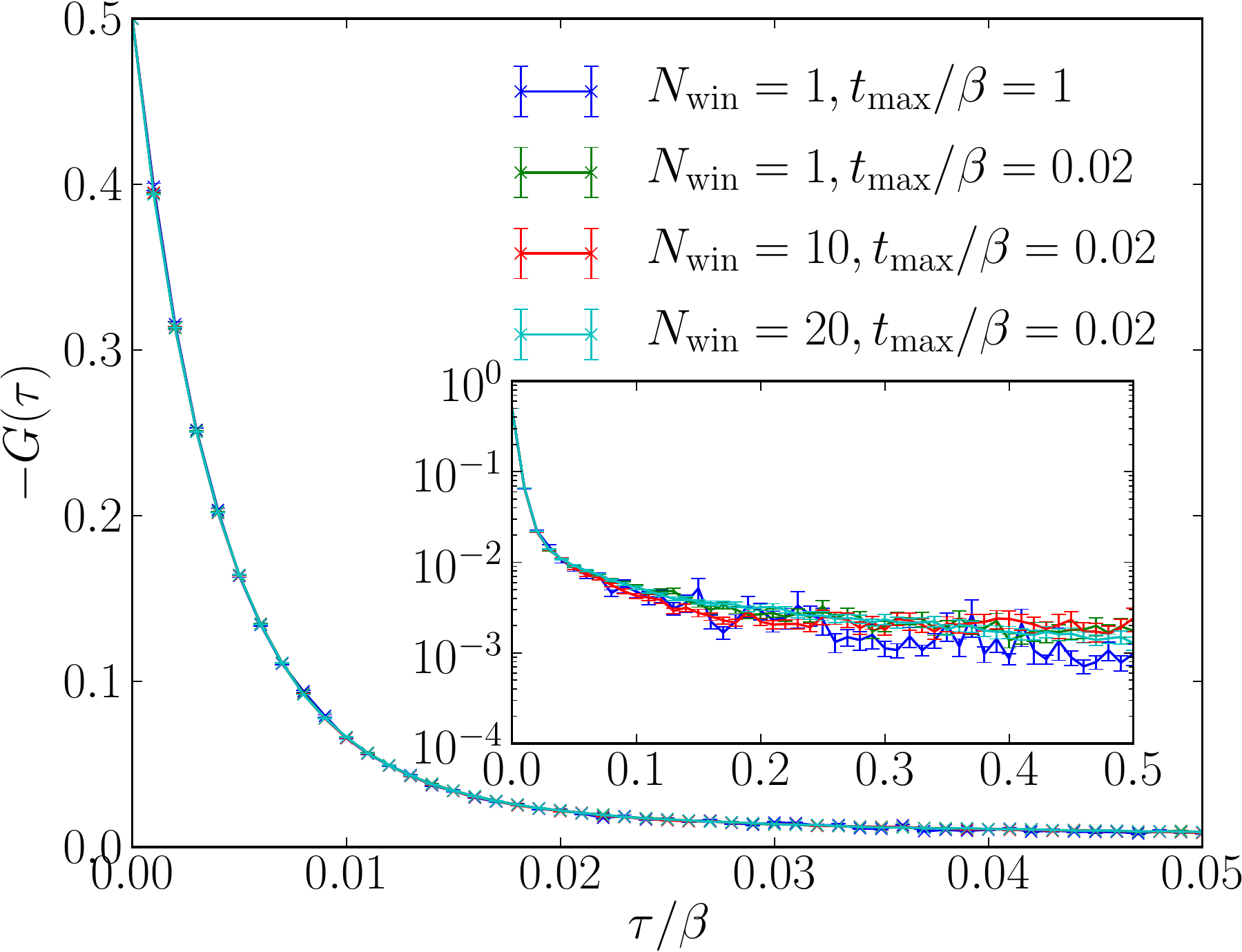}
 \caption{
 $G(\tau)$ computed at $U=6$ and $\beta=50$. The inset shows a log-scale plot.
 }
 \label{fig:gtau}
\end{figure}
In Fig.~\ref{fig:autocorr}(a), we show the distribution of 
the time difference between pairs of operators successfully removed or inserted in the Monte Carlo sampling performed with $t_\mathrm{max}/\beta=1$ and $N_\mathrm{win}=1$.
As expected, we see that the accepted updates are local in imaginary time.
The distribution decreases exponentially for large time differences.
We found that the range $\tau_\mathrm{pair}/\beta\le 0.02$ accounts for almost 94 \% of the successful updates. 
Considering the condition in Eq.~(\ref{eq:lower-bound}), we take $t_\mathrm{max}/\beta=0.02$ and $N_\mathrm{win}\le 20$.
Since $N_\mathrm{exp}\simeq 5.6$, the window contains on average 5.6 operators for $N_\mathrm{win}=20$.

In Fig.~\ref{fig:gtau}, we show the Green's function $G(\tau)$ computed using 
$t_\mathrm{max}/\beta=0.02$ for different values of $\tau_\mathrm{win}$.
We also present data obtained for $t_\mathrm{max}/\beta=1$ for comparison.
All the data shown are consistent within error bars, indicating our algorithm works correctly.
However, we found that $G(\tau)$ for $N_\mathrm{win}=1$ and $t_\mathrm{max}=1$ is systematically smaller than the others.
This may be because the autocorrelation time is too long for the MC simulation to be thermalized.

In Fig.~\ref{fig:autocorr}(b), we show the $N_\mathrm{win}$ dependence of the timing per MC step for $t_\mathrm{max}/\beta=0.02$.
As expected, the timing decreases linearly with the window size $\tau_\mathrm{win}$.
The estimated autocorrelation time is shown in Figs.~\ref{fig:autocorr}(c) in units of MC steps.
For $N_\mathrm{win}=1$, the autocorrelation time is shorter by one order of magnitude for $t_\mathrm{max}/\beta=0.02$ compared to that for $t_\mathrm{max}/\beta=1$.
This is consistent with the increase in the acceptance rate from 0.022 to 0.34 by introducing the cutoff.
Now, we discuss the $N_\mathrm{win}$ dependence. 
Around $\tau\simeq \beta/2$, the autocorrelation time is not affected badly by introducing the window, consistent with the above argument.
Although the autocorrelation is affected around $\tau=0.01$, the increase is considerably smaller than the reduction in the CPU time. 

Figure~\ref{fig:autocorr}(d) shows the autocorrelation time in units of seconds.
It is clearly seen that the autocorrelation becomes shorter in the entire $\tau$ region as $N_\mathrm{win}$ increases up to $N_\mathrm{win}=20$.
The improvement is as much as two orders of magnitude from the most naive approach 
($N_\mathrm{win}=1$ and $t_\mathrm{max}/\beta=1$)
to 
the best case ($N_\mathrm{win}=20$ and $t_\mathrm{max}/\beta=0.02$).

\subsubsection{Temperature and $U$ dependence}
\begin{figure}[!]
 \centering
 \includegraphics[width=.45\textwidth,clip]{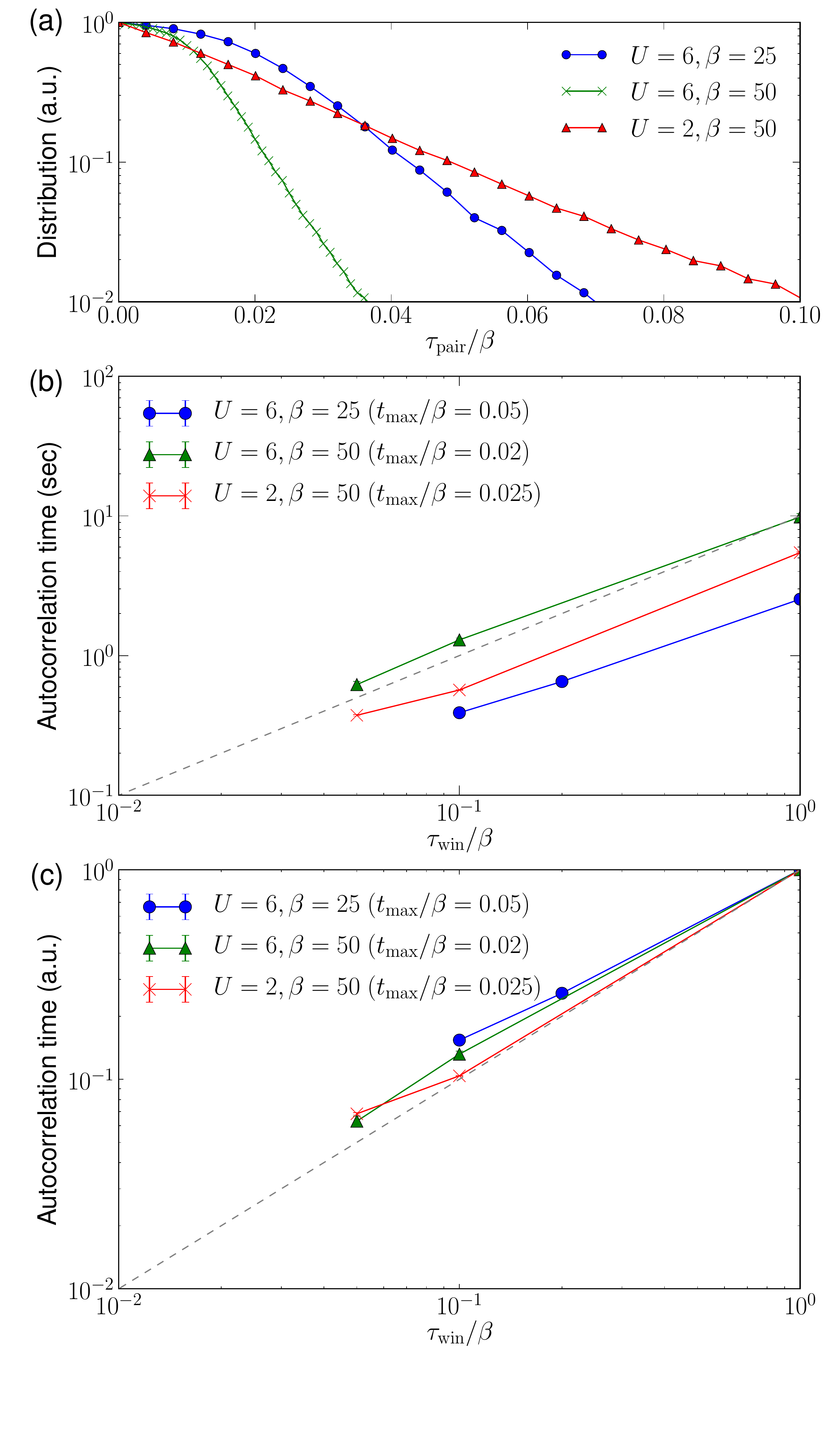}
 \caption{
 (a) Distribution of the time difference of a pair of operators successfully removed or inserted in the MC sampling.
 (b)/(c) $\tau_\mathrm{win}$ dependence of the autocorrelation time in CPU time.
 The autocorrelation time is averaged over the interval of $0<\tau<\beta$.
 In (c), the data are normalized by the timings for $\tau_\mathrm{win}/\beta=1$.
 }
 \label{fig:autocorr-scaling}
\end{figure}

Figure~\ref{fig:autocorr-scaling}(a) shows the distribution function of the length of successfully inserted and removed pairs of operators for different values of $\beta$ and $U$.
The weakly correlated metallic region corresponds to $U\lesssim 2$.
First, we discuss the temperature dependence for $U=6$.
Comparing the data for $\beta=25$ and $\beta=50$, one can see that the distribution becomes more localized at low temperatures.
This may be because the hybridization function $\Delta(\tau)$ decays more rapidly with $\tau$ similarly to $G(\tau)$ at low temperatures.
This result indicates that our improved MC sampling works even better at low temperatures.
On the other hand, although the distribution becomes broader at smaller $U$,
the distribution still decays exponentially at long distances.
In Figs.~\ref{fig:autocorr-scaling}(b) and (c), we plot the $\tau_\mathrm{win}$ dependence of the autocorrelation time averaged over the interval $0<\tau<\beta$.
We took $t_\mathrm{max}=0.05$, 0.02 and 0.025 for $(U=6, \beta=25)$, $(U=6, \beta=50)$ and $(U=2, \beta=50)$.
It is clearly seen that the autocorrelation time scales linearly with $\tau_\mathrm{win}$ down to the lowest $\tau_\mathrm{win}$ for all parameter sets.
This indicates the robustness of the sliding window approach.

\subsection{Discussion and future perspectives}\label{sec:smpl-discussion}
A simple way to choose the window size is to measure the distribution function of the distance between successfully inserted or removed pairs of operators during the thermalization process.
Then, one can choose a reasonable cutoff $t_\mathrm{max}$ such that most of the distribution, say 95\%, is contained within the cutoff. 
The window size $\tau_\mathrm{win}=\beta/N_\mathrm{win}$ is then given by the minimum size that satisfies the lower bound given in Eq.~(\ref{eq:lower-bound}).

Further improvement of the efficiency may be possible by
using the heat-bath algorithm or a better algorithm~\cite{Suwa-Todo-PRL} where we propose several candidates at each update.
This allows to increase the acceptance rate and reduce the autocorrelation time.

There are other kinds of local updates with acceptance rates higher than inserting/removing pairs of operators.
Examples include shifting an operator on the imaginary time axis or swapping two nearest neighboring operators.
Introducing such efficient updates helps in practical calculations.

\section{Summary}\label{sec:summary}
In this paper, we discussed two complementary approches based on the hybridization-expansion continuous-time Monte Carlo method for multi-orbital systems.
First, we proposed the combine the Krylov approach with the MPS/MPO representation of states and operators. 
We found that highly accurate results can be obtained by using bond dimensions considerably smaller than the dimension of the whole Hilbert space.
Based on a scaling analysis, we showed that the performance becomes superior to the conventional method for quantum impurity problems involving more than 12 orbitals.

Second, we proposed an improved Monte Carlo sampling algorightm for the hybridization expansion Monte Carlo method.
Detailed benchmark tests were carried out for a 5-orbital impurity model.
We showed that the new algorithm works robustly for a broad range of on-site repulsions and temperatures.
In particular, we confirmed that the ``sliding window" approach works particularly efficiently at low temperatures,
and we expect that it will be useful in the study of phenomena emerging at low temperatures. 
The sampling scheme is easy to implement in existing Monte Carlo codes,
and applies to any variant of the hybridization expansion method.

\appendix

\ack
We thank Iztok Pizorn on the discussion on matrix product states. 
We also thank Jakub Imriska, Hidemaro Suwa, Hugo Strand, Lei Wang, and Li Huang for useful comments on the improved Monte Carlo sampling.
HS and PW acknowledge support from the DFG via FOR 1346 and SNF Grant 200021E-149122. 
This project was supported by ERC grant SIMCOFE.
Simulations were performed using the ALPS libraries.~\cite{alps2}

\bibliographystyle{iopart-num}
\bibliography{ref}

\appendix

\section{MPO for a model with uniform all-to-all interactions}\label{sec:mpo-compression}
Let us consider a Hamiltonian with uniform all-to-all interactions:
\begin{eqnarray}
  \mathcal{H} &=& \sum_{n=1}^{N_\mathrm{op}} \sum_{i\ge 1, j\ge 2, i<j}^{L} \hat{A}^{(n)}_i \hat{B}^{(n)}_j + \sum_{i=1}^L \hat{O}_i,\label{eq:ham-all-to-all}
\end{eqnarray}
where $\hat{A}^{(n)}_i$ and $\hat{B}^{(n)}_j$ are operators acting on the local Hilbert spaces on sites $i$ and $j$, respectively.
$\hat{O}_i$ is an operator acting on site $i$.
A compressed MPO can be explicitly constructed for this kind of model with all-to-all uniform interactions.

The Hamiltonian in Eq.~(\ref{eq:ham-all-to-all}) may be written in the form 
\begin{eqnarray}
  \mathcal{H}  &=& \boldsymbol{W}_1 \boldsymbol{W}_2 \cdots \boldsymbol{W}_L,\label{eq:mpo-ham-all-to-all}
\end{eqnarray}
where $\boldsymbol{W}_i$ is a matrix whose elements are operators acting on the local Hilbert space at site $i$.
The $\boldsymbol{W}_i$ are given as follows:
\begin{eqnarray}
  \boldsymbol{W}_1 &=& \left(
  \begin{array}{ccccc}
    \hat{I} &  \hat{A}^{(1)} & \cdots & \hat{A}^{(N_\mathrm{op})}        & \hat{O}
  \end{array} 
  \right),\\
  \boldsymbol{W}_i &=& \left(
  \begin{array}{ccccc}
    \hat{I} & \hat{A}^{(1)} & \cdots & \hat{A}^{(N_\mathrm{op})} & \hat{O}\\
    0        & \hat{I} & 0 & 0 & \hat{B}^{(1)}\\
    0        & 0 & \ddots &  0 & \vdots\\
    0        & 0 & 0 & \hat{I} &  \hat{B}^{(N_\mathrm{op})}\\
    0        & 0 & 0 & 0 & \hat{I}\\
  \end{array} 
  \right),\\
\boldsymbol{W}_L &=& \left(
  \begin{array}{c}
\hat{O}\\
\hat{B}^{(1)}\\
\vdots\\
\hat{B}^{(N_\mathrm{op})}\\
\hat{I}\\
  \end{array} 
  \right),\label{eq:W}
\end{eqnarray}
where $1<i<L$ and $\hat{I}$ denotes the identity operator.
One can see that Eq.~(\ref{eq:mpo-ham-all-to-all}) is in the MPO form with bond dimension $N_\mathrm{op}+2$ when each element in $\boldsymbol{W}_i$ is regarded as a $4 \times 4$ matrix.

For the multi-orbital Hubbard model given in Eq.~(\ref{eq:imp}), one obtains an MPO of bond dimension eight by taking
\begin{eqnarray}
  \hat{A}^{(1)} &=& n_\uparrow,\\
  \hat{B}^{(1)} &=& (U^\prime-J) n_\uparrow + U^\prime n_\downarrow,\\
  \hat{A}^{(2)} &=& n_\downarrow,\\
  \hat{B}^{(2)} &=& (U^\prime-J) n_\downarrow + U^\prime n_\uparrow,\\
  \hat{A}^{(3)} &=& S^+~(\equiv c_\uparrow^\dagger c_\downarrow),\\
  \hat{B}^{(3)} &=& -J S^-~(\equiv -Jc_\downarrow^\dagger c_\uparrow),\\
  \hat{A}^{(4)} &=& S^-,\\
  \hat{B}^{(4)} &=& -J S^+,\\
  \hat{A}^{(5)} &=& D^+~(\equiv c_\uparrow^\dagger c_\downarrow^\dagger),\\
  \hat{B}^{(5)} &=& -J D^-~(\equiv -J c_\uparrow c_\downarrow),\\
  \hat{A}^{(6)} &=& D^-,\\
  \hat{B}^{(6)} &=& -J D^+,\\
  \hat{O} &=& U \hat{n}_\uparrow \hat{n}_\downarrow.
\end{eqnarray}
For the local Hilbert space spanned by $|0\rangle$, $\opcdag_{i\downarrow} |0\rangle$, $\opcdag_{i\uparrow} |0\rangle$, $\opcdag_{i\uparrow} \opcdag_{i\downarrow} |0\rangle$,
\begin{eqnarray}
  c^\dagger_\uparrow &=& 
  \left(
  \begin{array}{cccc}
  0 & 0 & 1 & 0\\
  0 & 0 & 0 & 1\\
  0 & 0 & 0 & 0\\
  0 & 0 & 0 & 0\\
  \end{array}
  \right),\\
  c^\dagger_\downarrow &=& 
  \left(
  \begin{array}{cccc}
  0 & 1 & 0 & 0\\
  0 & 0 & 0 & 0\\
  0 & 0 & 0 & -1\\
  0 & 0 & 0 & 0\\
  \end{array}
  \right),\\
  n_\uparrow &=& c^\dagger_\uparrow c_\uparrow,\\
  n_\downarrow &=& c^\dagger_\downarrow c_\downarrow.
\end{eqnarray}

\section{MPO for general interactions}\label{sec:mpo-general-int}
In this Appendix, we show how the MPS formalism is extended to general interactions.
Let us begin by showing the MPO representation of annihilation and creation operators.
In the operator representation, they look like 
\begin{eqnarray}
  && \hat{f}_1 \otimes \hat{f}_2 \otimes \cdots \otimes \hat{f}_{i-1} \otimes \hat{O}_i\otimes \hat{I}_{i+1} \cdots \otimes \hat{I}_L,
\end{eqnarray}
with the site index explicitly shown.
We omit the spin index for simplicity.
Here, $\hat{O}$ is the matrix representation of the annihilation or creation operators given in 
\ref{sec:mpo-compression}.
The operator $\hat{f}_i$ counts the number of particles, which is given by 
\begin{eqnarray}
  f &=& 
  \left(
  \begin{array}{cccc}
  1 & 0 & 0 & 0\\
  0 & -1 & 0 & 0\\
  0 & 0 & -1 & 0\\
  0 & 0 & 0 & 1\\
  \end{array}
  \right)
\end{eqnarray}
in the local basis introduced in 
\ref{sec:mpo-compression}.
Therefore, annihilation and creation operators are obviously represented by an MPO with bond dimension one.
Since the product of two MPO with bond dimension one has bond dimension one,
any product of annihilation and creation operators can be represented by an MPO with bond dimension one.
For example, a correlated hopping term $\hat{n}_1\hat{c}_2^\dagger\hat{c}_4$ reads
\begin{eqnarray}
  && \hat{n}_1 \otimes  \hat{c}^\dagger_2 \hat{f}_2\otimes \hat{f}_3 \otimes \hat{f}_4\hat{c}_4\otimes \hat{I}_5 \otimes \cdots.
\end{eqnarray}

The summation over the site index can be explicitly taken in a way similar to that in Appendix~\ref{sec:mpo-compression}.
Let us consider the sum of correlated hopping terms 
\begin{eqnarray}
  \sum_{i\neq j\neq k} \hat{n}_i \hat{c}^\dagger_j \hat{c}_k
\end{eqnarray}
as an example.
For simplicity, we restrict ourselves to the case $i<j<k$.
In this case, the sum is represented by the MPO with the following local tensors:
\begin{eqnarray}
  \boldsymbol{W}_1 &=& \left(
  \begin{array}{cccc}
    \hat{I} & \hat{n} & 0 & 0
  \end{array} 
  \right),\\
  \boldsymbol{W}_i &=& \left(
  \begin{array}{cccc}
    \hat{I}  & \hat{n} & 0        & 0 \\
    0        & \hat{I} & \hat{c}^\dagger\hat{f} & 0\\
    0        &      0  & \hat{f} & \hat{f}\hat{c} \\
    0        &      0  &      0  & \hat{I} \\
  \end{array} 
  \right)~(1<i<L),\\
\boldsymbol{W}_L &=& \left(
  \begin{array}{c}
    0\\
    0\\
    0\\
\hat{I}\\
  \end{array} 
  \right).\label{eq:W-chopping}
\end{eqnarray}

\begin{figure}[hb]
 \centering
 \includegraphics[width=.3\textwidth]{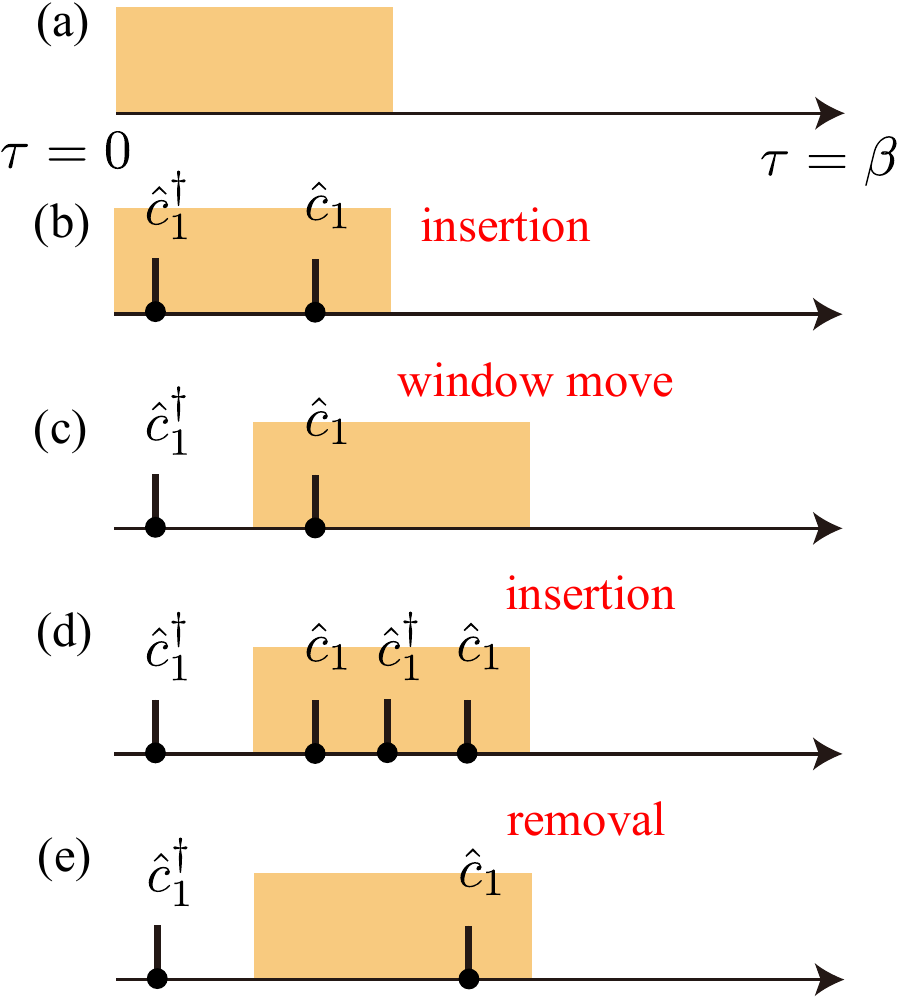}
 \caption{How to insert a pair of operators with an arbitrary time difference.}
 \label{fig:ergodicity}
\end{figure}

\section{Ergodicity of the sliding-window approach}\label{sec:ergodicity}
In this Appendix, we show that the MC sampling based on the sliding window approach is ergodic.
In particular, we show that a pair of operators with an arbitrary time difference can be inserted by repeated insertions and removals of pairs with a short time difference.
The procedure is illustrated in Fig.~\ref{fig:ergodicity}.
First, we insert a pair in the window as shown in Figs.~\ref{fig:ergodicity}(a) and (b).
Then, the window is moved to the next position [Fig.~\ref{fig:ergodicity}(c)].
As shown in Fig.~\ref{fig:ergodicity}(d), 
the distance between the operators can be increased by inserting a new pair and removing two operators in the middle because the two windows are overlapping each other.
By repeating this procedure, one can create a pair with an arbitrary time difference.
One can also remove any pair of operators, independent of the time difference, by reversing the above procedure.
Therefore, it is obvious that one can transform any configuration into any other configuration by inserting and removing pairs within the sliding window.

\section{Detailed Monte Carlo update procedure}\label{sec:MC-detail}
The local Monte Carlo update procedure has been described in Sec. II B of Ref.~\cite{Haule07}.
In this Appendix, we explain how this procedure is modified when the cutoff $t_\mathrm{max}$ and the sliding window are introduced. 

Let us consider an attempt to insert a pair of creation and annihilation operators of flavor $f$ at $\tau_\mathrm{c}$ and $\tau_\mathrm{a}$.
More specifically, we first choose $\tau_\mathrm{c}$ randomly and uniformly in the window.
Then, $\tau_\mathrm{a}$ is choosen randomly and uniformly in the window under the constraint $|\tau_\mathrm{c}-\tau_\mathrm{a}|\le t_\mathrm{max}$.
The reverse process of this update is removing one of operator pair of flavor $f$ whose length is equal or less than $t_\mathrm{max}$.

We first discuss the case without a cutoff $t_\mathrm{max}$.
The window is located on the interval $[\tau_\mathrm{win}^\mathrm{min},\tau_\mathrm{win}^\mathrm{max}]$ with $\tau_\mathrm{win}=\tau_\mathrm{win}^\mathrm{max}-\tau_\mathrm{win}^\mathrm{min}$.
The weights of the original and new configurations are denoted by $w_\mathrm{org}$ and $w_\mathrm{new}$, respectively.
The probability to accept this insertion is 
\begin{eqnarray}
  P&=&\mathrm{min}\left[1, \left|\frac{w_\mathrm{new}}{w_\mathrm{org}}\right| \frac{\tau_\mathrm{win}^2}{N_\mathrm{pair}^{t_\mathrm{max}}}\right],
\end{eqnarray}
where $\tau_\mathrm{win}=\tau_\mathrm{win}^\mathrm{max}-\tau_\mathrm{win}^\mathrm{min}$ is the size of the window and
$N_\mathrm{pair}$ is the number of operator pairs of flavor $f$ in the window after the insertion.

By introducing a cutoff $t_\mathrm{max}$, the probability is changed to 
\begin{eqnarray}
  P&=&\mathrm{min}\left[1, \left|\frac{w_\mathrm{new}}{w_\mathrm{org}}\right| \frac{\tau_\mathrm{win}\Delta \tau_\mathrm{a}}{N_\mathrm{pair}^{t_\mathrm{max}}}\right],
\end{eqnarray}
where 
\begin{eqnarray}
  \Delta \tau_\mathrm{a} &=& \mathrm{min}(\tau_\mathrm{c}+t_\mathrm{max},\tau^\mathrm{max}_\mathrm{win})
  - \mathrm{max}(\tau_\mathrm{c}-t_\mathrm{max},\tau^\mathrm{min}_\mathrm{win})\hspace{5mm}
\end{eqnarray}
and $N_\mathrm{pair}^{t_\mathrm{max}}$ is the number of operator pairs of flavor $f$ whose length is equal or less than $t_\mathrm{max}$ in the window after the insertion.

The probability to accept an attempt to remove a kink at $\tau_\mathrm{c}$ and $\tau_\mathrm{a}$ is correspondingly given by
\begin{eqnarray}
  P&=&\mathrm{min}\left[1, \left|\frac{w_\mathrm{new}}{w_\mathrm{old}}\right| \frac{N_\mathrm{pair}^{t_\mathrm{max}}}{\tau_\mathrm{win}\Delta \tau_\mathrm{a}}\right],
\end{eqnarray}
where $N_\mathrm{pair}^{t_\mathrm{max}}$ is the number of operator pairs of flavor $f$ for the original configuration.

\end{document}